\documentstyle[aps,prb,twocolumn]{revtex}

\def\sfont#1{\centerline{\sc{#1}}}

\begin{document}
\title{\hfill {\small ITP Preprint Number NSF-ITP-95-03} \vspace{10pt} \\
Problems With the Vortex-Boson Mapping in 1+1 Dimensions}

\author{Leon Balents}
\address{Institute for Theoretical Physics, University of California,
Santa Barbara, CA 93106-4030}
\author{Steven H.  Simon}
\address{Department of Physics, Harvard University, Cambridge, MA
  02138 \\ \vspace{.3in} \centerline{\begin{minipage}[t]{6in} \rm
Using the well known boson mapping, we relate the transverse magnetic
susceptibility of a system of flux vortices in 1+1 dimensions to an
appropriately defined conductivity of a one-dimensional boson system.
The tilt response for a system free of disorder is calculated
directly, and it is found that a subtle order of limits is required to
avoid deceptive results.\end{minipage}}}

\maketitle
\setcounter{section}{1}
\vspace{-.3in}
{\sfont{1. Introduction}}
\label{intro} \vspace{10pt}

A $d+1$ dimensional vortex system with correlated disorder can be
modeled as a $d$ dimensional time dependent boson system.\cite{Seung}\
This mapping has proved extremely useful for the study of vortex
behavior for systems with disorder correlated in one
direction.\cite{BGtheory}\  Such pinning may arise from artificially
created ion tracks, screw dislocations, and/or twin or grain
boundaries.\cite{Disorder}\ Unfortunately, difficulties in
characterizing the disorder lead to uncertainties in the
interpretation of experimental
measurements.\cite{VGexperiments,BGexperiments}\  In contrast, in the
1+1 dimensional system of flux vortices in large Josephson junctions,
disorder can be fabricated with great control.\cite{Itzler}\  In
particular, it is easy to create samples with disorder
appropriately correlated in one direction.  This freedom has
encouraged both the experimental\cite{Itzler}\ and
theoretical\cite{US,FGB,VK}\ study of these 1+1 dimensional systems.
Researchers have also found 1+1 dimensional vortex models to be a
useful simplification for understanding the physics of the more
complicated 2+1 dimensional systems.\cite{RGrefs,HwaBatrouni}\

In this paper, expanding on previous work,\cite{Seung,US}\ we discuss
features of the boson mapping for 1+1 dimensional systems.  The basic
mapping is briefly reviewed in section 2.  As might be expected, the
response of the vortex system to a transverse magnetic field (the tilt
response) is related to a conductivity of the boson system. As is well
known, one must be very careful about defining a conductivity in one
dimension.\cite{Fisher2}\ The relation between the vortex tilt
response and a boson conductivity is derived in section 3.  We will
see, however, that this result is extremely deceptive due to a subtle
order of limits.  To illustrate this problem, as well as the method of
treating it properly, we explicitly calculate the tilt response for a
system with no disorder in section 4.\vspace{10pt}

\setcounter{section}{2}
{\sfont{2. Boson Analogy}}
\vspace{10pt}

\label{BosonAnalogy}
We consider a system of vortices confined to the $x-y$ plane and
oriented along the $y$ axis, with disorder correlated in the
$y$-direction.  In this case, such the
pinning potential can be written as $U(x)$ and the free energy ${\cal
F_N}$ for a system of $N$ vortices is\cite{Seung,US}
\begin{eqnarray}
  & & {\cal F}_N = -N(H/H_{c1}-1) \epsilon L + \nonumber \\ & &
  \int_0^L dy \bigg\{ \sum_{n} \bigg[ {\tilde{\epsilon} \over 2}\left|
  {{dx_n(y)} \over {dy}} - h(x_n(y),y))\right|^2 +
  U(x_n(y)) \bigg] \nonumber \\ & & +
  \sum_{n,n'} V(|x_n(y)-x_{n'}(y)|) \bigg\},
\label{modelequation}
\end{eqnarray}
where $x_i(y)$ describes the path of the $i^{th}$ vortex, $H_{c1}$ is
the lower critical field, $h(x,y)=H_x(x,y)/H_y(x,y)$ is the local
slope of the applied magnetic field, $\tilde \epsilon$ is the vortex
stiffness, and $V(x)$ is the vortex-vortex interaction.

If periodic
boundary conditions are imposed on the system in the $y$ direction,
the partition function for the system of flux lines is given by
\begin{equation}
  \label{eq:partition}
  Z =  \sum_{N=0}^{\infty} \frac{1}{N!} \sum_P \int {\cal D} x_1(y)
  \cdots \int {\cal D} x_N(y) \, e^{-{\cal F}_N/k_{\rm B} T}
\end{equation}
where $y$ goes from $0$ to $L$.  The sum over $P$ is a sum over all
possible permutations of boundary conditions, such that for all $i$,
$x_i(0) =x_j(L)$ for some $j$.  As discussed in Ref.
\onlinecite{Seung}, in the limit of $h(y)=0$, this partition function
is identical to an imaginary time Feynman path integral for the grand
canonical partition function of a collection of bosons in 1+1
dimensions.  The $y$ coordinate for the flux line system is mapped to
the imaginary time coordinate of the boson system.  The chemical
potential of boson system is given by $\epsilon (H/H_{c1J}-1)$, and
the boson mass is given by the fluxline stiffness $\tilde \epsilon$.
The temperature $k_{\rm B} T$ of the vortex is mapped to Plank's
constant $\hbar$ for the boson system, and the length $L$ of the
vortex system is $\beta \hbar$ with $\beta$ the inverse temperature
($1/k_{\rm B} T_{bose}$) of the bose system.  The boson pair potential
is given by the flux line pair interaction $V(|x_i - x_j|)$, and the
$y$-correlated disorder potential $U(x_i)$ for the vortex system also
maps directly to the bose system.
\vspace{10pt}

\setcounter{section}{3}
{\sfont{3. Tilt Response}}
\label{TiltResponse2}
\vspace{10pt}

The above described mapping  is particularly useful in allowing
one to make contact with the large body of knowledge of interacting
boson systems.  To make this comparison more fruitful, we now derive a
relation between the tilt response of the vortex array and the
conductivity of the interacting bosons.  As will be discussed in an
explicit calculation in section 4, it is necessary
to keep careful track of the order of limits defining the dc response
to avoid deceptive results.

To probe the linear response of the flux line system, a small field in
the $x$ direction can be applied.  To avoid problems with flux lines
piling up at the $x$ boundaries of the system,\cite{Fisher2}\ the field
is applied over a finite width $0<x<W$, and we will take the limit $W
\rightarrow \infty$ at the end.  For simplicity, we will choose $H_x$
to be given in the form
\begin{equation}
  h(x,y)= \left\{ \begin{array}{lll} h(y) & \mbox{for} & 0 < x < W \\
  0 & & \mbox{otherwise}
  \end{array} \right.
\end{equation}
although it is a trivial generalization to include more complicated
variations of $h(x,y)$ in the $x$-direction.  The Fourier components
of the function $h(y)$ are defined by
\begin{equation}
  \tilde h(q) = \frac{1}{L}\int_0^L dy e^{iqy} h(y)
\end{equation}
where $q$ must be of the form $2 \pi n/L$ with $n$ an integer.  It
should be noted that since $h(y)$ is real, $\tilde h(q)=\tilde h(-q)^*$.  The
tilt
modulus $t(q)$ can now be defined as
\begin{equation}
  t(q;W) = \left. \frac{d \tilde \theta(q;W)}{d\tilde h(q)} \right|_{h=0}
\end{equation}
where $\tilde \theta(q;W)$ is the Fourier component of the local slope
\begin{equation}
  \tilde \theta(q;W) = \frac{1}{L} \int_0^L dy e^{iqy}   \theta(y;W)
\end{equation}
and $  \theta(y;W)$ is the spatially averaged slope of the flux vortices
in the range $0<x<W$.  More formally,
\begin{equation}
    \theta(y;W) = \frac{1}{W} \int_0^W dx \, s(x,y)
  \label{eq:thetadef2}
\end{equation}
where $s(x,y)$ is the local slope density
\begin{equation}
  s(x,y) = \sum_i \frac{dx_i}{dy} \delta(x-x_i(y)).
\end{equation}
The expectation of the slope $\tilde \theta(q;W)$ is given by
differentiating the log of the partition function
\begin{equation}
\left< \theta(q;W) \right> = \frac{k_{\rm B}T}{WL \tilde \epsilon} \frac{d \ln
  Z}{d\tilde h(-q)}.
\end{equation}
Thus, the tilt response is the second derivative, which can be written
as
\begin{eqnarray}
  t(q;W) &=& \frac{k_{\rm B}T}{W L \tilde \epsilon} \frac{d^2 \ln
    Z}{d\tilde h(q)d\tilde h(-q)} \\
  &=& \label{eq:tint1}
\frac{W \tilde \epsilon}{k_{\rm B}T}
 \left< \int_0^L dy e^{-iqy}   \theta(0;W)   \theta(y;W) \right>
\end{eqnarray}
where we have used the fact that $\left< \tilde \theta(q;W) \right>$
is zero in the limit of zero field, as well as the $y$-translational
invariance of the system.  In the limit $L \rightarrow \infty$ this
expression becomes analogous to the expression for the conductivity of
the bose system at zero temperature\cite{Fisher2,FisherLee}\ and finite
frequency $\omega$ where the external field is applied over the range
$0<x<W$:
\begin{equation}
\label{eq:MPAsigma}
\sigma(\omega;W) =  \frac{e^2}{\pi \hbar \omega}
\int_{-\infty}^{\infty} d\tau e^{-i\omega\tau} \left< T_\tau   J(\tau)
  J(0) \right>
\end{equation}
where $T_\tau$ is the time ordering operator, and $  J(\tau)$ is
the current at imaginary time $\tau$ averaged over $0<x<W$,
\begin{equation} \label{eq:Jint}
    J(\tau) = \frac{1}{W} \int_0^W dx j(x,\tau)
\end{equation}
with the local current density
\begin{equation}
  j(x,\tau) = \sum_i \frac{dx_i}{d\tau} \delta(x-x_i(y)).
\end{equation}
Thus with $q$ taking the place of $\omega$, the formal relation is
\begin{equation}
  t(q;W) = q \sigma(q;W) \frac{W\tilde \epsilon}{k_{\rm B} T}
 \frac{\pi \hbar}{e^2}.
\label{tcrelation}
\end{equation}

\vspace{10pt}
\setcounter{section}{4} {\sfont{4. Calculation of the Tilt
    Response:}} {\sfont{Clean Limit --- Luttinger Liquid}}
\label{TiltResponse}
\vspace{10pt}

Eq.\ref{tcrelation}\ is appealing, but can lead to
drastically incorrect results if applied naively.  In an interacting
one--dimensional quantum system, we expect a finite dc conductivity.
Taking the limit $q \rightarrow 0$, Eq.\ref{tcrelation}\ seems to
imply a vanishing tilt response to a uniform field.  It is intuitively
clear, however, that the vortices should rotate uniformly to
accommodate the transverse field if there are no pinning defects
present.  To resolve this apparent paradox, we now calculate the
conductivity more carefully for general $q$ and $W$ (with $W$ much
larger than the mean vortex/boson spacing but still finite).

If there is no disorder in the system, the ground state is a periodic
array of vortices.  Fluctuations of this array are described by a
dimensionless displacement field $u(x,y)$, where $x_n(y) = \ell(n +
u(n\ell,y)/2\pi)$, with $\ell$ the mean vortex separation.  The free
energy can then be written as
\begin{equation}
  \label{eq:luttinger}
  \frac{{\cal F}}{k_{\rm B}T} = \int_{-\infty}^{-\infty} dx \int_{-L/2}^{L/2}
  dy \left[ \frac{K_x}{2} \left| \frac{du}{dx} \right|^2 +
  \frac{K_y}{2} \left| \frac{du}{dy} \right|^2 \right]
\end{equation}
Here we have shifted the zero value of $y$.

Within the boson mapping, Eq.\ref{eq:luttinger}\ can be interpreted as
the bosonized action for the quantum particles.  Generic values of
$K_x$ and $K_y$ actually correspond to the non-Fermi-liquid
``Luttinger liquid'' state induced by interactions in one dimension.
It is satisfying that this highly non--trivial behavior of the quantum
system is encoded in our simple displacement field description.  The
values of the coefficients $K_x$ and $K_y$ are dependent on the
precise form of the interaction $V$ as well as on the average spacing
$\ell$.  Standard methods\cite{CIT}\ can be used to evaluate these
constants in various limits.\cite{BigNote}

By transforming into Fourier space, the free energy becomes a sum (ie,
an integral) of uncoupled harmonic oscillator modes.  The
equipartition theorem then yields
\begin{equation}
  \left<\tilde u_{q_x,q_y} \tilde u_{q_x',q_y'}\right> =
  \frac{(2\pi)^2 \delta(q_x+q_x') \delta(q_y+q_y')}{K_x q_x^2 + K_y q_y^2},
\end{equation}
where $\tilde u_{q_x,q_y}$ is the Fourier transform of $u(x,y)$.
With the current ($j(x,\tau)$) or local slope density given in terms
of the displacement field as $s(x,y) = du/dy$, the slope averaged
from $0$ to $W$ (see Eq.\ref{eq:thetadef2}) can be written as
\begin{equation}
    \theta(y;W) = \int_0^W  \frac{dx}{W}\int
  \frac{dq_x dq_y}{(2\pi)^2}   e^{i(x q_x +y q_y)} iq_y \tilde u_{q_x,q_y}
\end{equation}
Performing   the   $x$     integration  and    substituting       into
Eq.\ref{eq:tint1}\ yields the tilt response
\begin{equation}
  t(q;W) =  \frac{2 W \tilde \epsilon q^2}{k_{\rm B} T (2\pi)} \int dq_x
\frac{(1
    - \cos(q_x W))}{(q_x W)^2 \left[K_y q^2 + K_x q_x^2\right]}
\end{equation}
By adding a small piece $\delta^2$ to the $(q_x W)^2$ term of the
denominator, the integral can be performed by contour integration and
the limit $\delta \rightarrow 0$ can be taken at the end to give the
result
\begin{equation}
  t(q;W) =  \frac{2 \tilde \epsilon}{K_y k_{\rm B} T} f
\left(qW\sqrt{K_y/K_x}\right)
\label{cleantilt}
\end{equation}
with
\begin{equation}
  f(x) =  \left[ 1 - \frac{1-e^{-x}}{x} \right].
\label{fdef}
\end{equation}
As noted above, this calculation can easily be generalized to the case
where we are concerned with the response to a field $h(x,y)$ that is
an arbitrary function of $x$.

Eq.\ref{cleantilt}\ is an appealing physical result, and resolves the
``paradox'' described at the beginning of this section.  To define
the conductivity of the quantum system, we must take $q\rightarrow 0$
for finite $W$, to prevent difficulties with equilibration and
transport across the ends of the sample.  Then, using
$\lim_{x\rightarrow 0} f(x) = x/2$ and Eq.\ref{tcrelation}, the dc
conductivity is finite.  In the vortex system, however, it is clear
that for finite $W$ the fluxons cannot tilt, since they would have to
pile up at the boundaries to match the untilted flux lines outside the
field region.  Taking therefore $W\rightarrow\infty$ first, the
behavior $f(x) \rightarrow 1$ in this limit gives a finite $t(q=0)$.

\vspace{20pt}
\setcounter{section}{5}
{\sfont{5. Conclusion}}
\label{Conclusion}
\vspace{10pt}

We have shown that a surprising subtlety in the order of limits arises
in the calculation of tilt response and conductivity even for the
simple pure problem discussed above.  The origin of the difficulty
lies in boundary effects, which are generally ignored in applications
of the vortex--boson analogy.  It is natural that transport properties
such as the conductivity are sensitive to changes at the boundary, due
to the constraint of charge conservation.  One might conjecture that
in the {\sl localized} Bose glass phase, the order of limits may become
unimportant.  This possibility, and the extent to which such problems
manifest themselves in other types of correlation functions remain
interesting open questions.\\

\vspace{10pt}

{\sfont{Acknowledgments}}

This research was supported by the National
Science Foundation under Grant No. PHY89--04035 at the ITP, and
National Science Foundation Grant DMR-91-15491 at Harvard.

\vspace{10pt}

\begin{references}

\bibitem{Seung}  D. R. Nelson, Phys. Rev. Lett. {\bf 60}, 1973 (1988);
D. R. Nelson and H. S. Seung, Phys. Rev. B {\bf 39}, 9153 (1989).




\bibitem{BGtheory} D. R. Nelson and V. M. Vinokur, Phys. Rev.
B{\bf 48}, 13060 (1993).

\bibitem{Disorder} See, e.g. S. Flesher et. al., Phys. Rev. B{\bf 47},
14448 (1993); M. Hawley et. al., Science {\bf 251}, 1587 (1991); and
Ref.\onlinecite{VGexperiments}.

\bibitem{VGexperiments} R. H. Koch, V. Foglietti, W. J. Gallagher, G.
Koren, A. Gupta, and M. P. A. Fisher, Phys. Rev. Lett. {\bf 63}, 1511
(1989); P. L. Gammel, L. F. Schneemener, and D. J. Bishop, Phys. Rev.
Lett. {\bf 66}, 953 (1991).

\bibitem{BGexperiments} L. Civale, A. D. Marwick, T. K. Worthington,
M. A. Kirk, J. R. Thompson, L. Krusin-Elbaum, Y. Sum, J. R. Clem and
F. Holtzberg, Phys. Rev.  Lett. {\bf 67}, 648 (1991); M. Leghissa,
L. A. Gurevich, M. Kraus, G.  Saemann--Ischenko, and L. Ya. Vinnikov,
Phys. Rev. B {\bf 48}, 1341 (1993).



\bibitem{Itzler} M. A. Itzler and M. Tinkham, Phys. Rev. B{\bf 51},
  435 (1994); unpublished; to appear in {\it IEEE Trans. Appl.
    Supercond.}, (1994).


\bibitem{US} Leon Balents and Steven H. Simon, Phys. Rev. B, In Press.


\bibitem{FGB} R. Fehrenbacher, V. B. Geshkenbein, and G. Blatter,
Phys. Rev. B{\bf 45}, 5450 (1992).

\bibitem{VK} V. M. Vinokur and A. E. Koshelev, Zh. Eksp. Teor. Fiz.
{\bf 97}, 976 (1990) [Sov. Phys. JETP {\bf 70}, 547 (1990)].


\bibitem{RGrefs} T. Natterman, I. Lyuksyutov, and M. Schwartz,
  Europhys. Lett. {\bf 16}, 295 (1991); J. Toner, Phys. Rev. Lett.
  {\bf 67}, 2537 (1991); Y.-C. Tsai and Y. Shapir, Phys. Rev. Lett.
  {\bf 69}, 1773 (1992).

\bibitem{HwaBatrouni} G. G. Batrouni and T. Hwa, Phys. Rev. Lett. {\bf
72}, 4133 (1994).


\bibitem{Fisher2} D. S. Fisher and P. A. Lee, Phys. Rev. B {\bf 23},
  6851 (1981).


\bibitem{CIT} See, e.g. S. N. Coppersmith, D. S. Fisher, B. I.
Halperin, P. A. Lee, and W. F. Brinkman, Phys. Rev. B {\bf 25}, 349
(1982).

\bibitem{BigNote} For the physically relevant case of the Josephson
  junctions described in Ref. \onlinecite{US}, near the lower critical
  field ($H/H_{c1J} - 1 \ll 1$) where the fluxon density is low,
  standard methods\cite{CIT}\ give (with notation from Ref.
  \onlinecite{US}) $K_x = ck_{\rm B}T/(4\pi\tilde{\epsilon}\ell)$ and
  $K_y = \tilde{\epsilon}\ell/(4\pi k_{\rm B}T c)$, with $c$ a
  non--universal order one constant.  In the high magnetic field limit
  ($H \gg \phi_0/d \lambda_J$), these methods are not appropriate
  since the vortices overlap and become poorly defined.  None the
  less, it can be shown that the free energy defined in
  Eq.\ref{eq:luttinger} still applies, with $K_x \approx K_y \approx
  \epsilon_J/k_{\rm B}T$.


\bibitem{FisherLee} C. L. Kane and M. P. A. Fisher, Phys. Rev. Lett.
  {\bf 68}, 1220 (1992).






\end{references}
\end{document}